\documentclass[10pt,twocolumn,letterpaper]{article}

\usepackage[pagenumbers]{cvpr} 

\usepackage{graphicx}
\usepackage{amsmath}
\usepackage{amssymb}
\usepackage{booktabs}

\usepackage[pagebackref,breaklinks,colorlinks]{hyperref}

\usepackage[capitalize]{cleveref}
\crefname{section}{Sec.}{Secs.}
\Crefname{section}{Section}{Sections}
\Crefname{table}{Table}{Tables}
\crefname{table}{Tab.}{Tabs.}

\def\cvprPaperID{7193} 
\def\confName{CVPR}
\def\confYear{2022}

\begin{document}

\title{Online Meta Adaptation for Variable-Rate Learned Image Compression}

\author{Wei Jiang, Wei Wang, Songnan Li, Shan Liu\\
Tencent America LLC.\\
Palo Alto, CA\\
{\tt\small \{vwjiang,rickweiwang,sunnysnli,shanl\}@tencent.com}
}
\maketitle

\begin{abstract}
This work addresses two major issues of end-to-end learned image compression (LIC) based on deep neural networks: variable-rate learning where separate networks are required to generate compressed images with varying qualities, and the train-test mismatch between differentiable approximate quantization and true hard quantization. We introduce an online meta-learning (OML) setting for LIC, which combines ideas from meta learning and online learning in the conditional variational auto-encoder (CVAE) framework. By treating the conditional variables as meta parameters and treating the generated conditional features as meta priors, the desired reconstruction can be controlled by the meta parameters to accommodate compression with variable qualities. The online learning framework is used to update the meta parameters so that the conditional reconstruction is adaptively tuned for the current image. Through the OML mechanism, the meta parameters can be effectively updated through SGD. The conditional reconstruction is directly based on the quantized latent representation in the decoder network, and therefore helps to bridge the gap between the training estimation and true quantized latent distribution. Experiments demonstrate that our OML approach can be flexibly applied to different state-of-the-art LIC methods to achieve additional performance improvements with little computation and transmission overhead.
\end{abstract}

\section{Introduction}
\label{sec:intro}

Lossy image compression has been a decades-long research topic \cite{JPEG,JPEG2000,HEVC,VVC}, which converts images into as few bits as possible for efficient transmission and storage and then reconstructs the images from the transmitted bitstream. Motivated by the success of deep neural networks (DNNs) in a variety of computer vision tasks, end-to-end learned image compression (LIC) has been actively explored in recent years \cite{Hyperprior2017,ScaledHyperprior2018,ContextHyperprior2018,AdaptiveContext2018,ContextICLR2019,Cheng2020}. 

Great success has been achieved by DNN-based LIC. However, two major open issues still require further study: 1) the difference between the differentiable entropy model for rate estimation in training and the true distribution of the quantized latent representation at test time, and 2) the variable rate problem where multiple model instances are needed to meet compression requirements of different rate-distortion (RD) tradeoffs. 
Specifically, the DNN-based LIC is usually formulated as a joint RD optimization problem using a variational auto-encoder (VAE) architecture. The encoder/decoder network is trained to minimize the empirical RD loss of training images using a soft differentiable approximate quantization. For a test image, the true distribution of its hard quantized latent representation can be quite different from the estimated one. That is, in addition to the common generalization issue of all learning methods caused by the difference between training and test data, LIC models face another gap caused by the difference between the training entropy estimation for soft approximation and the true distribution of the hard quantized latent. To minimize this gap, since the pioneer work of \cite{Hyperprior2017}, several methods have been developed to improve the entropy model for accurate rate estimation, such as the scaled hyperprior \cite{ScaledHyperprior2018}, the conditional adaptive context \cite{AdaptiveContext2018}, the joint autoregressive and hyperprior, the context-adaptive joint autoregressive and hyperprior \cite{ContextICLR2019}, and the discretized GMM with attention \cite{Cheng2020}. However, the mismatch originates from the deterministic quantization error that is not truly random, which remains as a major source of the suboptimal RD performance for LIC \cite{quantization2020}.

On the other hand, for RD optimization, a hyperparameter $\lambda$ controls the tradeoff between the compression rate and reconstruction quality. Once trained for one tradeoff $\lambda$, the model usually can not perform well for another tradeoff $\lambda$. Therefore, one model instance usually needs to be trained for each tradeoff $\lambda$ for variable-rate LIC, which can be very expensive and inefficient. Some methods have been developed to tackle this variable rate issue, such as using specially designed generalized octave concolution/transposed-convolution and training loss in the encoder/decoder networks \cite{OctaveConv2021}, using a model-agnostic multi-task prune-and-grow strategy to share parameters among the encoder/decoder networks of multiple compression rates \cite{PnG2021}, or using a conditional variational auto-encoder (CVAE) to reconstruct images of different compression rates based on the rate control Lagrange multiplier as conditional variables \cite{conditionalVAE2019,modulateVAE2019}.

In this paper, we address both above issues by formulating the DNN-based LIC problem into an online meta-learning (OML) setting. The online learning mechanism is used to bridge the gap between the training soft approximate quantization and the true hard quantization at test time. The meta-learning mechanism is used to control reconstruction with different RD tradeoffs. 

From the perspective of machine learning, on the encoder side, LIC is a task with ground-truth target in the test stage. This makes LIC well suited for online deep learning (ODL), which can alleviate the problem caused by the mismatch of training entropy model and hard quantization at test time. However, we are reluctant to modify the model parameters based on any individual test image, which can be highly unstable with poor generalization \cite{ODLIJCAI2018}. In this work, instead of modifying the DNN model parameters, we use the OML framework to directly update the meta parameters for each test image through SGD. Such meta parameters enables effective adaptation in decoder for better reconstruction tuned to each particular image. 

Specifically, we develop our OML based on the CVAE architecture. We treat the conditional variables as meta parameters, treat the network module of generating conditional features as the meta prediction network, and treat the generated conditional features as meta priors. The conditional features modulate with the intermediate features of the decoder layers to compute meta-conditional features. In this formulation, the meta parameters control the desired reconstruction quality so that one model instance can accommodate reconstruction of different RD tradeoffs. Direct SGD is used to effectively update the meta parameters so that the decoder can adapt its reconstruction according to the current test image.  

Our contributions can be summarized as follows:
\begin{itemize}
\item We formulate the LIC problem into an OML setting to address both the variable-rate learning issue and the gap between the training soft approximate quantization and the true hard quantization at test time. 
\item Different from general OML for life-long model learning \cite{OMLICML2019,OMLLCLR2019}, we online update the conditional meta parameters instead of model parameters, for the sake of both system stability and transmission efficiency. The updated meta parameters control the conditional reconstruction based on quantized latent representation directly, which tunes the reconstructed image according to the current need. 
\item The proposed OML-based LIC framework is based on the CVAE architecture and can be applied to various underlying VAE models.
\end{itemize}
We evaluate our algorithm over the JPEG-AI benchmark dataset provided in the MMSP 2020 challenge \cite{MMSP2020}. Experimental results demonstrate the effectiveness of the proposed method in improving the state-of-the-art underlying LIC models \cite{Cheng2020}.

\section{Related Work}
\label{sec:relatedwork}

\subsection{Learned image compression}

Many recent LIC methods take the VAE architecture, where image compression is formulated into the joint RD optimization using variational inference \cite{Hyperprior2017}. Since additive uniform noise is used during training to approximate the test-time quantization, most studies focused on improving the entropy model to reduce the mismatch between the estimated distribution and the true distribution of quantized latent representation \cite{ScaledHyperprior2018,ContextHyperprior2018,AdaptiveContext2018,3DContext2020,Cheng2020}. However, such a mismatch originates from the deterministic quantization error that is not truly random, the gap between the soft quantization approximation and the true hard quantization can not be eliminated. 

\subsection{Variable-rate LIC}

When trained end-to-end, the encoder/decoder network is optimized for a specific RD tradeoff, which will not work well for other tradeoffs. As a result, variable-rate LIC is typically quite hard, since one network is trained per compression rate. To achieve variable-rate flexibility similar to the modern codecs, both RNN-based and VAE-based solutions have been proposed to improve the model design or training strategies. 

The convolutional/deconvolutional LSTM was used in \cite{LSTMVariable2016,LSTMVariable2017} for incremental reconstruction. The RNN-based architecture requires progressive encoding/decoding, and the many iterations needed for high-quality reconstruction can be hard to use in practice. 

Under the VAE framework, to pursue high RD performance for variable-rate compression, the work of \cite{OctaveConv2021} designed  the generalized octave convolution/transposed-convolution in the residue blocks of the encoder/decoder network. However, it is non-trivial to apply the  method to general model structures. To pursue general applicability, a model-agnostic multi-task learning strategy was proposed in \cite{PnG2021} to share parameters among encoders/decoders of multiple compression rates. However, the success of the method relies on appropriate sharing structures and sharing ratios, as well as skillful training using the prune-and-grow strategy. In \cite{conditionalVAE2019,modulateVAE2019}, the CVAE architecture was used to control compression rates, where the Lagrange multiplier for rate control was given as conditional variables to the network. Conditional inference was performed through generating conditional features that were concatenated with common features \cite{conditionalVAE2019} or modulated with common features \cite{modulateVAE2019}.

We develop our OML based on the CVAE architecture, in pursuit of a balance between general applicability and RD performance, since CVAE can accommodate a range of VAE-based LIC model structures. In addition, CVAE can be interpreted as a meta-learning model structure, as will be shown in Section \ref{sec:ourmethod}. SGD can be used to effectively update the meta parameters for our online learning.

\subsection{Online Learning}

Online learning aims to improve generalization of machine learning models, \textit{i.e.}, to alleviate the problem caused by different training and test data distributions.
The problem of LIC is well suited for online learning, since the target is to encode and recover the input image itself, and the encoder has the ground-truth input at test time. In this paper, we use online learning to bridge not only the common gap between the training and test data distributions, but also the gap between the training soft approximate quantization and true hard quantization.

Most online learning methods focus on online updating the learned models \cite{OLJMLR2011,OLNIPS2013}, and their performance with DNNs for online deep learnig (ODL) is quite limited \cite{ODLIJCAI2018}. This is because the highly complex DNN models need to be trained with batch-based methods using mini-batches and multiple passes over the training data. Updating model parameters on the per-sample basis can be highly unstable. 

Here we take a different strategy for ODL. Instead of online updating the encoder/decoder networks, we directly alter the meta parameters that control the reconstruction process based on quantized latent representation. Direct SGD is effectively used to update the meta parameters according to the desired target loss, so that the reconstruction is tailored for the current data adaptively. 

\subsection{Meta Learning}

Meta-learning aims to learn from the experience of a set of machine learning tasks so that learning of a new task can be fast. In the context of LIC, if we treat compression with each target RD tradeoff as a task, by observing training tasks of multiple compression rates, meta learning enables fast generalization to a new test compression rate, which is analogous to variable-rate LIC.

Assume that tasks are drawn from a task distribution, and a set of training tasks with their corresponding datasets are observed. Then a meta-learning algorithm tries to learn a task-general prior over the model parameters, and such prior knowledge can be applied to a new task to speedup its learning. Among various meta-learning methods \cite{metanips2017,metaiclr2018,MAML2017,ICLR2019}, the gradient-based model-agnostic meta-learning (MAML) \cite{MAML2017,ICLR2019} has been successfully used in various applications, \textit{e.g.}, reinforcement learning \cite{MAMLRL2019}, image super-resolution \cite{MAMLSR2020}, HDR image reconstruction \cite{MAMLHDR2021}, \textit{etc}. 

For the scenario of continual learning, where the task distribution is not fixed but changing overtime, the online meta-learning (OML) framework has been developed \cite{OMLICML2019,OMLLCLR2019}, where the MAML meta-training with direct SGD is performed online during a task sequence to update the parameters of the task model. 

Motivated by the OML mechanism, we perform SGD to change the conditional meta parameters that control the reconstruction process in decoder. Compared with updating model parameters, another benefit besides stability is that only a few updated conditional meta parameters need to be transmitted to the decoder, in comparison to the large number of model updates.

\subsection{Substitutional LIC}
\label{sec:relatedsubstitution}

Another work highly related to to ours is the substitutional LIC method proposed in \cite{SubstitutionNIC}. The original input image is replace by a substutional image that outperforms the original one for a new target, \textit{e.g.}, a new target metric or a new target compression tradeoff. Similar to our method, direct SGD is used to iteratively update the substitute image from the original input, without modifying the underlying encoder/decoder networks. Different from \cite{SubstitutionNIC}, where updates are conducted in the input image domain as a preprocessing module using training entropy estimation, we present meta online adaptation in the latent domain based on reconstruction from true quantized latent representation, to alleviate the problem of mismatch between the soft approximate quantization and true hard quantization, under the framework of variable-rate CVAE.

Another benefit of online adaptation over true quantized latent is that we avoid multiple iterative passes of context computation. For example, the autoregressive context model has good RD performance but is slow in computation, due to the sequential scan order. To alleviate this issue, some recent methods have been developed to enable parallelization, {\textit e.g.}, though two-pass checkboard context calculation \cite{checkerboardcontext2021}.  Here we can avoid multiple passes of context computation by conducting online adaptation in the decoder reconstruction network.

\section{Online Meta-LIC}
\label{sec:ourmethod}

A typical LIC architecture consists of an encoder $f_{\Phi}(x)$, a decoder $g_{\Theta}(z)$, and a quantizer $q(\cdot)$, where $x$ is an input image and $z\!=\!q(f_{\Phi}(x))$ is a quantized latent representation from the quantizer. Lossless arithmetic coding is generally used to further generate a compressed bitstream from the quantized representation z for transmission. 

Since the deterministic quantization is non-differentiable with regard to network parameter $\Phi, \Theta$, the additive uniform noise is generally used to optimize an approximated differentiable RD loss comprising of: 
\begin{eqnarray}
\mathbf{R}_{\Phi} = E_{p(x)p_{\Phi}(z|x)}[-\log_2 p_{\Phi}(z)]\nonumber\\
\mathbf{D}_{\Phi,\Theta} = E_{p(x)p_{\Phi}(z|x)}\left[||x-g_{\Theta}(z)||^2_2\right],\nonumber 
\end{eqnarray}
where $p(x)$ is the probability density function of all natural images. Since the continuous density $p_{\Phi}(z)$ is 
intractable to compute, a differentiable tractable density $q_{\Theta}(z)$ is used to approximate $p_{\Phi}(z)$, and the RD optimization turns to:
\begin{eqnarray}
\min_{\Phi,\Theta} E_{p(x)p_{\Phi}(z|x)} \left[\lambda||x-g_{\Theta}(z)||^2_2-\log_2 q_{\Theta}(z)\right],\label{eqn:RDoriginal}
\end{eqnarray}
where $\lambda$ is a hyperparameter that controls the optimization of the network parameters to trade off between reconstruction quality against compression rate. Therefore, for each target value of $\lambda$, a set of parameters $\Phi, \Theta$ needs to be trained for the corresponding optimization of Eqn.(\ref{eqn:RDoriginal}), which is highly inefficient in practice.

\subsection{Variable-rate LIC}

Using the CVAE architecture, the variable-rate LIC conducts VAE image compression conditioned on the compression rates controlled by $\lambda$ . The following RD formulation is generally used:
\begin{eqnarray}
\min_{\Phi,\Theta} E_{p(x)p_{\Phi}(z|x,\lambda)} \left[\lambda||x-g_{\Theta}(z,\lambda)||^2_2-\log_2q_{\Theta}(z|\lambda)\right],\label{eqn:CVAERD}
\end{eqnarray}
where $\lambda\!\in\!\Lambda$, and $\Lambda$ is a set of Lagrange multiplier values. In \cite{conditionalVAE2019}, $\Lambda$ contains a pre-defined finite set of values, and a one hot vector of length $|\Lambda|$ is generated as the input condition. In our OML setting, we need to assign arbitrary continuous values to $\Lambda$. Therefore, we use the conditional feature modulation framework similar to \cite{modulateVAE2019}, where $\Lambda$ can take any arbitrary value. 

\begin{figure}
\centering
\begin{subfigure}[b]{0.4\linewidth}
   \includegraphics[width=1\linewidth]{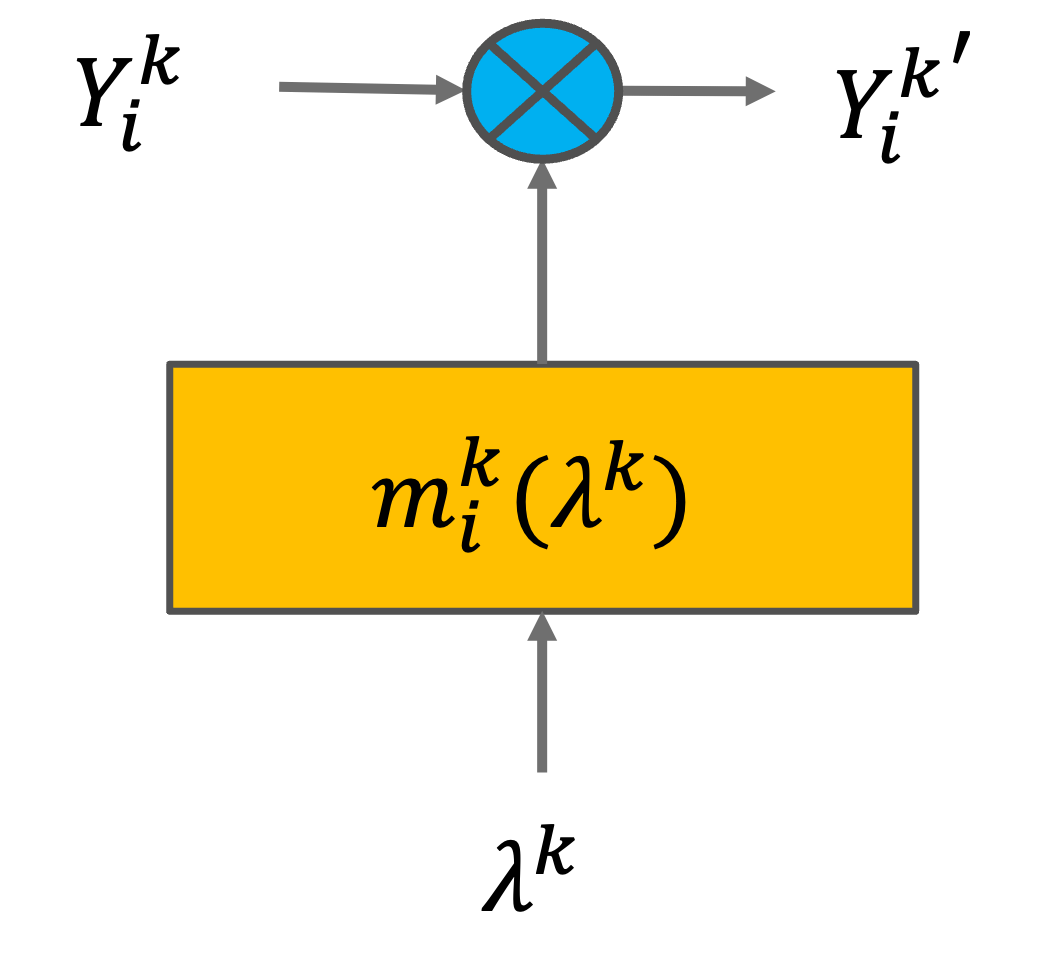}\vspace{-.5em}
\end{subfigure}
   \caption{Conditional feature modulation}\label{fig:CVAEmod}
\end{figure}

Specifically, the CVAE inference is implemented via a feature modulation operation illustrated in Figure \ref{fig:CVAEmod}:
\begin{eqnarray}
{Y^k_i}^{'} = s^k_i(\lambda^k)Y^k_i,
\end{eqnarray}
where $Y^k_i$ and ${Y^k_i}^{'}$ are the 2D feature maps of the $i$-th  input channel of the $k$-th layer before and after modulation, respectively.  $s^k_i(\lambda^k)$ is the channel-wise scaling factor, which depends on $\lambda^k$. $\lambda^k$ is the quality-control tradeoff condition for the $k$-th layer, and 
$\lambda^1\!=\!\lambda^2\!=\!\ldots\!=\!\lambda$ given the target tradeoff $\lambda$ of the current compression. $s^k_i(\lambda^k)$ can be computed as:
\begin{eqnarray}
s^k_i = \text{softplus}(m^k_i(\lambda^k|\Psi^k_i))\nonumber
\end{eqnarray}
where $m^k_i(\lambda^k|\Psi^k_i)$ is a nonlinear function that maps the tradeoff $\lambda^k$ to the scaling factor of the $i$-th  channel and $\text{softplus}(x)\!=\!\log(1\!+\!e^x)$, and $\Psi^k_i$ contains the model parameters of $m^k_i$.

\subsection{Variable-rate meta-LIC}

Assume that the tasks of LIC with different $\lambda$s are drawn from a task distribution, $\cal{T}$. At meta-training time, we observe $M$ tasks with $\lambda_1,\ldots,\lambda_M$. At test time, we have a new task with an arbitrary $\lambda_t$. By learning from the training tasks, meta-learning-based LIC aims to optimize the RD loss for $\lambda_t$, without regular large-scale training for $\lambda_t$. 
Let $\Psi\!=\!\{\Psi^k_i\}$ include all the parameters for the conditional modulation. Let $L(D_j,\lambda_j,\Psi)$ represent the average loss on the dataset $D_j$ for RD tradeoff $\lambda_j$. The MAML method \cite{MAML2017} learns an initial set of parameters $\Psi$ based on all the training tasks, by solving the optimization problem:
\begin{eqnarray}
\Psi^*=\text{argmin}_{\Psi}\sum_{j=1}^M L(D_j,\lambda_j,\Psi-\alpha\Delta\hat{L}_j(\Psi,\lambda_j)),
\end{eqnarray}
where $\Delta\hat{L}_j(\Psi,\lambda_j)$ is the inner gradient computed based on a small mini-batch of data $D_j$, and $\alpha$ is the step size. Then at meta-test time, $L(D_t,\Psi,\lambda_t)$ can be minimized by performing a few steps of gradient descent from $\Psi$ using new task data $D_t$. 

In the context of online LIC, the current task is to compress the test image $x$, and we have $D_t\!=\!x$ and $|D_t|\!=\!1$. 

\subsection{Online variable-rate meta-LIC}

Now we want to use the online learning framework to bridge the gap between the training soft approximate entropy estimation and the actual distribution of the quantized latent representation. However, Updating the model parameters $\Psi$ based on a single test datum can be highly unstable. Besides, the model updates need to be transferred to the decoder for reconstruction, which can be quite expensive. 

In this work, for online meta-LIC, at meta-test time, instead of updating model parameters $\Psi$, we minimize $L(D_t,\lambda_t,\Psi)$ by performing gradient descent over the tradeoff conditional factor $\lambda_t$:
\begin{eqnarray}
\lambda_t^{'} = \lambda_t-\gamma\Delta L(D_t,\lambda_t,\Psi)\label{eqn:lambdaupdate}
\end{eqnarray}
That is, the direct SGD is used to find a better tradeoff condition $\lambda_t^*\!=\!\{{\lambda^k_t}^*\}$ than the original $\lambda_t$, so that a better RD loss $L(D_t,\lambda_t^*,\Psi)$ can be obtained. Note that different from the original variable-rate LIC where $\lambda_t$ is the same across all modulated layers, the online meta LIC usully has a different ${\lambda^k_t}^*$ for each $k$-th conditional modulated layer learned through online SGD. 

The intuitive rationale behind this approach is that through meta training, the relationship between the conditional hyperparameters $\lambda$ and the loss $L(D,\lambda,\Psi)$ has been established by the CVAE decoder network. Therefore, when we fix input data $D_t$ and network $\Psi$, we can finetune $\lambda$ to reduce $L(D_t,\lambda,\Psi)$ tailored for the current input $D_t$. 

\section{Implementation Details}

\begin{figure*}
\centering
\begin{subfigure}[b]{0.9\textwidth}
   \includegraphics[width=1\linewidth]{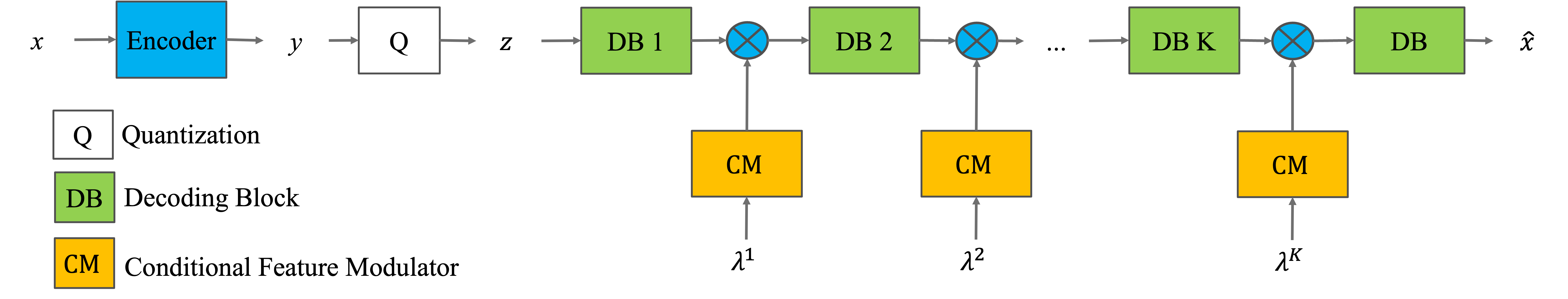}\vspace{-.5em}
   \caption{Variable-rate meta-LIC}
   \label{fig:metavariablerateLIC} 
\end{subfigure}\vspace{.5em}\vfill
\begin{subfigure}[b]{0.4\textwidth}\centering
   \includegraphics[width=0.7\linewidth]{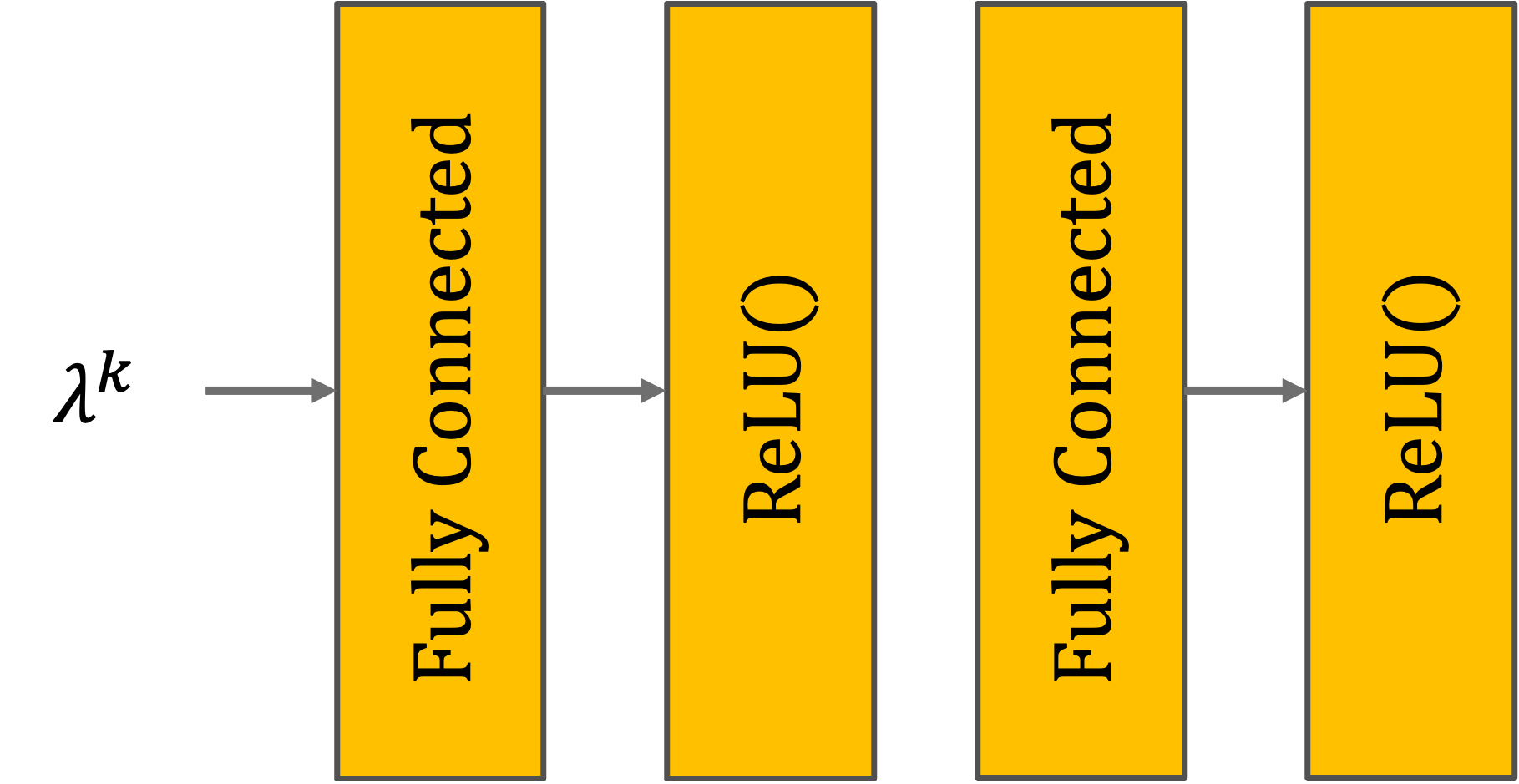}
   \caption{Conditional feature modulator (CM) network structure}
   \label{fig:modulatornetwork}
\end{subfigure}
\begin{subfigure}[b]{0.25\textwidth}\centering
   \includegraphics[width=0.65\linewidth]{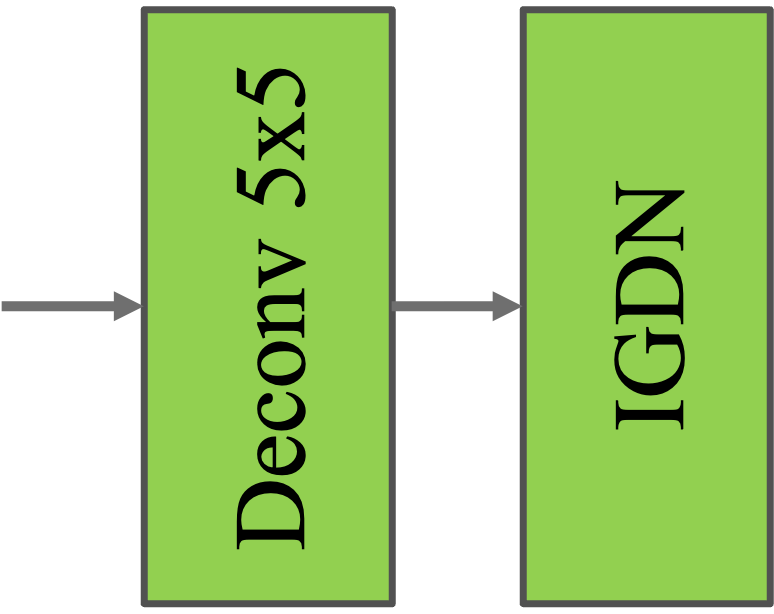}
   \caption{Decoding block (DB) for \cite{ContextHyperprior2018}}
   \label{fig:decodingblockhyper}
\end{subfigure}
\begin{subfigure}[b]{0.25\textwidth}
   \includegraphics[width=0.65\linewidth]{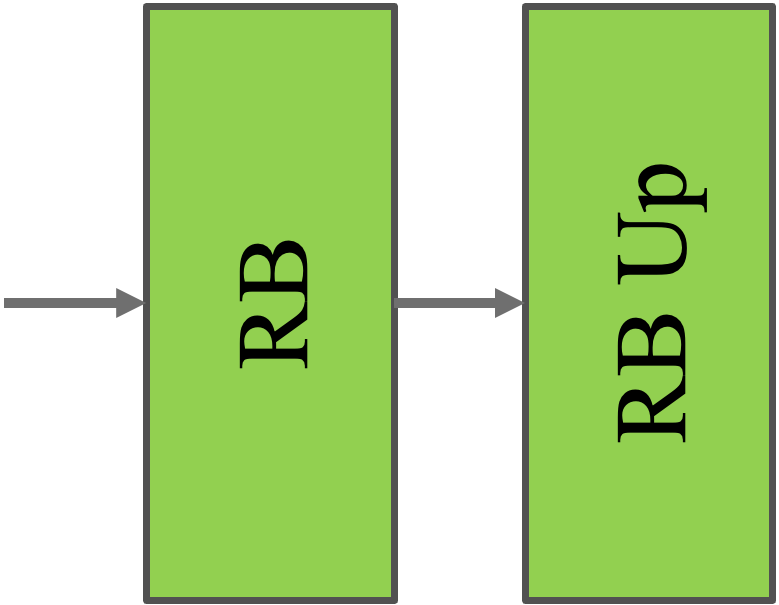}
   \caption{Decoding block (DB) for \cite{Cheng2020}}
   \label{fig:decodingblockcheng}
\end{subfigure}\vfill
\begin{subfigure}[b]{0.4\textwidth}\centering
   \includegraphics[width=0.75\linewidth]{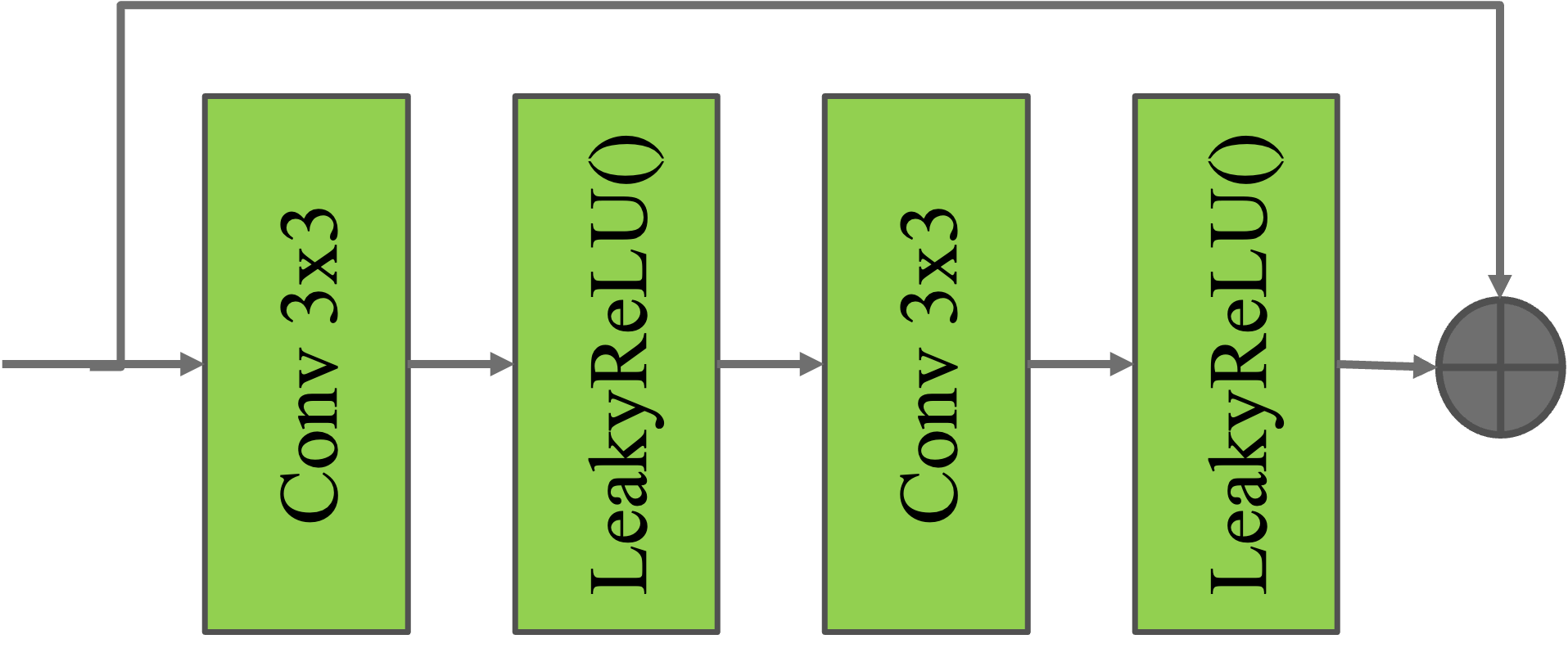}
   \caption{Residual block (RB) for \cite{Cheng2020}}
\end{subfigure}
\begin{subfigure}[b]{0.4\textwidth}\centering
   \includegraphics[width=0.75\linewidth]{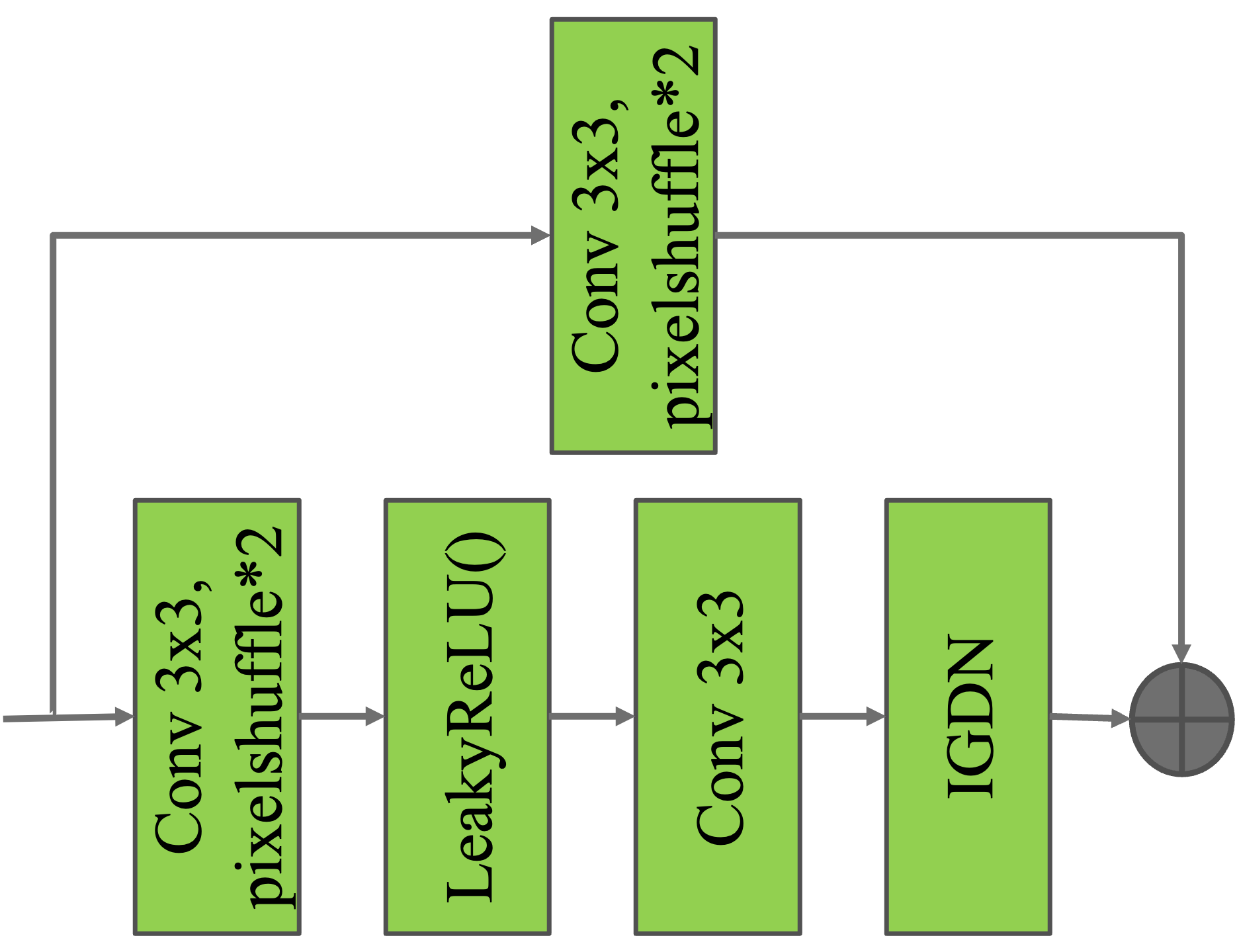}
   \caption{Residual upsampling block (RB Up) for \cite{Cheng2020}}
\end{subfigure}
\caption{Architecture of the proposed variable-rate meta-LIC method. The conditional feature modulators are applied to the decoder to perform channel-wise modulation over intermediate decoding features after the decoding blocks.}\label{fig:metaonlineLIC}
\end{figure*}

Figure \ref{fig:metaonlineLIC} describes the model architecture of our approach. The online meta learning aims to update the conditional parameters that control the reconstruction based on the quantized latent representation $z$ in Eqn.~(\ref{eqn:CVAERD}) directly. Therefore, it is unnecessary to enforce a variable-rate architecture for the encoder. In other words, for a base LIC encoder/decoder network, we only need to add the conditional modulation networks to modulate the intermediate features after the decoding blocks, with or without making changes to the encoders, as described in Figure~\ref{fig:metavariablerateLIC}.

As discussed in Section \ref{sec:relatedsubstitution}, another benefit of performing online adaptation in the reconstruction network is the computation, where the online SGD avoids multiple passes of expensive context calculation. 

This framework also gives the end users the flexibility to change the reconstructed image based on their different requirements. For instance, an encoder can send a universal quantized latent representation $z$ to user, together with a set of conditional tradeoff parameters $\{\lambda^{*}_{t}\}$, each $\lambda^{*}_{t}$ optimized for a target RD loss $\lambda_t\mathbf{D}\!+\!\mathbf{R}$, \textit{e.g.}, with different distortion metrics, different tradeoff values \textit{etc}. The end user can choose the optimal reconstruction according to  specific needs on the decoder side.

In detail, the conditional feature modulator network takes a similar structure to \cite{modulateVAE2019}, as described in Figure \ref{fig:modulatornetwork}. The conditional tradeoff parameter $\lambda^k$ for the $k$-th modulator is passed through two fully connected layers to generate a vector of length $N^k$, where $N^k$ is the number of feature channels for the intermediate feature to modulate.  

The conditional feature modulator can be flexibly applied to various underlying LIC decoders. In this work, we use the Minnen2018 method \cite{ContextHyperprior2018} and the Cheng2020 method \cite{Cheng2020} as examples, which are generally used as state-of-the-art baseline LIC methods in the LIC community \cite{MMSP2020}. Their decoding block structures are given in Figures \ref{fig:decodingblockhyper} and \ref{fig:decodingblockcheng}, respectively. The last decoding block usually comprises of a basic (convolution or residual) block followed by an upsampling layer (\textit{e.g.}, deconvolution or pixel shuffle layer).

\begin{figure*}
\centering
\begin{subfigure}[b]{0.8\linewidth}
   \includegraphics[width=1\linewidth]{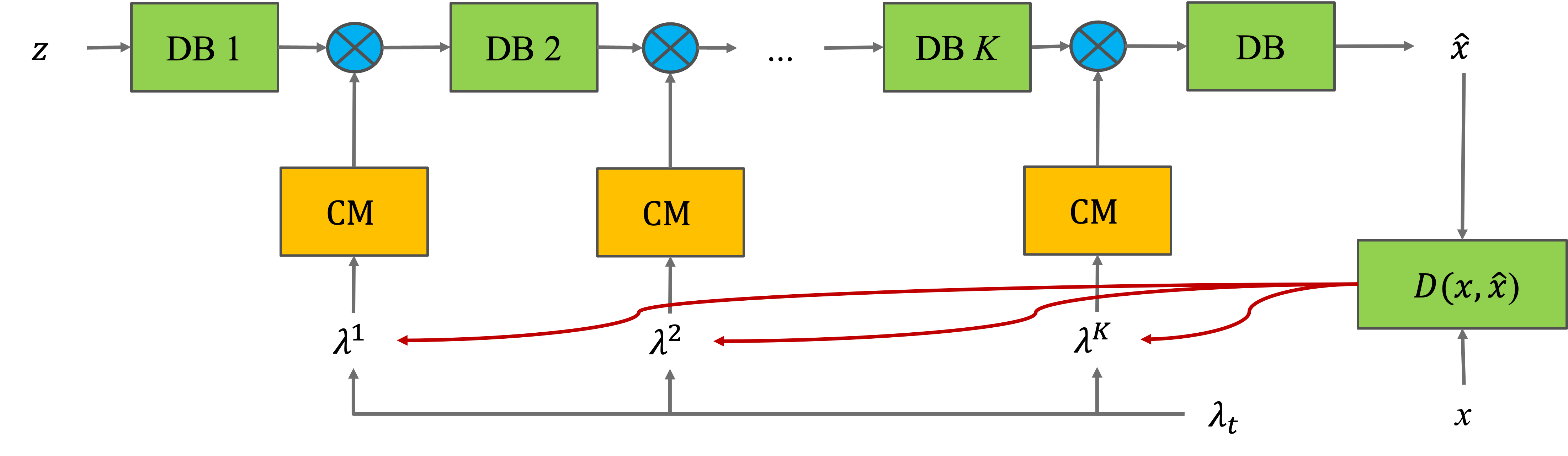}\vspace{-.5em}
\end{subfigure}
   \caption{Online meta-update of conditional tradeoff parameters through SGD.}\label{fig:onlineupdate}
\end{figure*}

Figure \ref{fig:onlineupdate} shows the process of online adaptation on the encoder side. Specifically, given an input image $x$ to compress, the encoder first computes the quantized latent representation $z$ based on the target tradeoff $\lambda_t$. The quantized latent $z$ is further encoded through lossless entropy coding and sent to the decoder. At the same time, through the 
CVAE decoder with conditional inputs $\lambda_t^1,\ldots,\lambda_t^K$, the reconstructed image $\hat{x}$ can be recovered, and the distortion $D(x,\hat{x})$ can be computed by the target distortion metric (\textit{e.g.}, PSNR or MSSSIM \cite{MSSSIM}). Initially we set $\lambda_t^1\!=\!\lambda_t^2\!=\!\ldots\!=\!\lambda_t^K\!=\!=\lambda_t$.

Since we do not change the quantized latent or the encoded bitstream of the quantized latent, the actual compression bpp (bits per pixel) remains the same during out OML process. Gradients can be computed based on the distortion loss $D(x,\hat{x})$, which can be backpropagated to update the conditional tradeoff parameters $\lambda_t^1,\ldots,\lambda_t^K$. After $n$ online iterations, the best performing updates ${\lambda_t^1}^*,\ldots,{\lambda_t^K}^*$ are sent to the decoder, which gives a better reconstructed image $\hat{x}$ with less distortion than the original $\lambda_t^1\!=\!\lambda_t^2\!=\!\ldots\!=\!\lambda_t^K\!=\!\lambda_t$.

\section{Experiments}

We conduct experiments using the JPEG-AI benchmark dataset provided by the MMSP 2020 challenge \cite{MMSP2020}. Our implementation is based on the CompressAI PyTorch package for LIC \cite{compressAI}. The JPEG-AI dataset comprises of 5264, 350, and 40 training, validation, and test images, respectively, with resolutions ranging from 256×256 to 8K. This is one of the latest and largest benchmark datasets for LIC research and standardization activities. We use both the Minnen2018 method \cite{ContextHyperprior2018} and the Cheng2020 method \cite{Cheng2020} as the base encoder/decoder networks, where the conditional modulator networks are added into the corresponding decoder as described in Figure \ref{fig:metavariablerateLIC}.

For the Minnen2018 method, CompressAI provides 8 pre-trained models corresponding to 8 compression quality levels with $\lambda_1,\ldots,\lambda_8$ as below:
\begin{eqnarray*}
\lambda_1=0.0018; \lambda_2=0.0035; \lambda_3=0,0067; \lambda_4=0.0130; \\ \lambda_5=0.0250; \lambda_6=0.0483;
\lambda_7=0.0932; \lambda_8=0.1800
\end{eqnarray*}
For the Cheng2020 method, CompressAI has 6 pre-trained models corresponding to the first 6 compression quality levels corresponding to $\lambda_1,\ldots,\lambda_6$. All pre-trained models aim to optimize PSNR, where distortion in the RD loss is measured by MSE. For Cheng2020, quality $\lambda_1,\lambda_2,\lambda_3$ have the same model architecture, and quality $\lambda_4,\lambda_5,\lambda_6$ have another same model architecture with more feature channels. Similarly for Minnen2018, quality $\lambda_1,\lambda_2,\lambda_3,\lambda_4$ have the same model architecture, and quality $\lambda_5,\lambda_6,\lambda_7,\lambda_8$ have another same model architecture with more feature channels. 

\subsection{Variable-rate base model}

The pre-trained models from CompressAI are used as base models, where the pre-trained encoders are directly used without any modification. The CVAE framework described in Figure \ref{fig:metavariablerateLIC} is added into the corresponding decoders to retrain the variable-rate decoder with the JPEG-AI training data through MAML meta-training \cite{MAML2017}.  That is, for Minnen2018, we have two meta-trained decoders corresponding to  pre-trained encoders for quality $\lambda_1,\lambda_2,\lambda_3,\lambda_4$ and $\lambda_5,\lambda_6,\lambda_7,\lambda_8$, respectively. For Cheng2020, we have two meta-trained decoders corresponding to pre-trained encoders for quality $\lambda_1,\lambda_2,\lambda_3$ and $\lambda_4,\lambda_5,\lambda_6$, respectively. The training is based on $256\!\times\!256$ patches randomly cropped from training images. 

\subsection{Online meta adaptation}

For each given test image $x$, it is divided into  $512\!\times\!512$ patches to feed into the LIC network, where online adaptation is conducted for each patch individually. $512\!\times\!512$ is the maximum patch size, \textit{i.e.}, small images and boundary patches will directly go through inference. For each patch, the original encoder computes the quantized latent representation, whose compression bpp stays unchanged for our meta online adaptation. Also, we send $K$ updated hyperparameters $\lambda^*_1,\ldots,\lambda^*_t$ as 16-bit float numbers to the decoder for each $512\!\times\!512$ patch. Since $K$ is a very small number ($K\!=\!4$ in our experiments), there is almost no additional transmission overhead (about 0.00025 bpp).

When $\lambda^k_t$ is updated through Eqn.~(\ref{eqn:lambdaupdate}), the updating step size $\gamma$ needs to be determined. We take a heuristic method to choose $\gamma$. For the first SGD iteration, we try several different step sizes (\textit{i.e.}, $\gamma\!=\!0.01,0.1,1,10,100,1000$) and select the best $\gamma^*$ with the minimum reconstruction distortion. Then we conduct $n$ iterations using $\gamma^*$, where at any iteration, if the distortion does not improve, we reduce $\gamma^*$ by half. Finally, the best performing $\lambda^k_t$ is recorded as ${\lambda^k_t}^*$. 

\subsection{Experimental results}

\begin{figure}
\centering
\begin{subfigure}[b]{\linewidth}
   \includegraphics[width=1\linewidth]{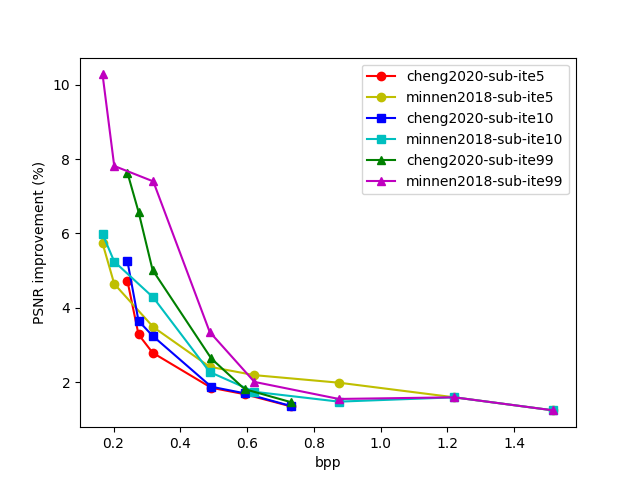}\vspace{-.5em}
   \caption{OML adaptation targeting at PSNR. }\label{fig:psnr}
   \end{subfigure}
\begin{subfigure}[b]{\linewidth}
   \includegraphics[width=1\linewidth]{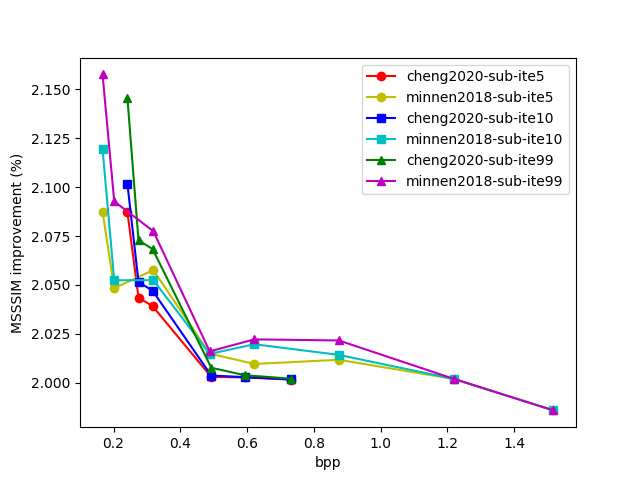}\vspace{-.5em}
      \caption{OML adaptation targeting at MSSSIM.}\label{fig:ssim}   
\end{subfigure}\vspace{-.5em}
   \caption{Performance improvements of OML over different base models. The MSSSIM is magnified as: $-10\log_{10}(1-\text{MSSSIM})$}  \vspace{-1em}
\end{figure}

Figure~\ref{fig:psnr} and Figure~\ref{fig:ssim} shows the performance improvements of OML over different baseline methods for different bpps, targeting at online adaptation for PSNR and MSSSIM, respectively. The results show that our OML method consistently improves the underlying variable-rate decoder for all measured qualities. In general, the OML adaptation helps the low-bitrate reconstruction more than the high-bitrate reconstruction. Also, the improvements over Minnen2018 are more than cheng2020. This is quite reasonable, since the low-bitrate latent representation corresponds to large reconstruction artifacts, which gives more room for OML to finetune the features. In comparison, high-bitrate latent is good for reconstruction itself, and can be less robust to online changes. Also, the better the underlying encoder/decoder in context modeling, the less the training-test gap for OML to help with. 

Note that it is within expectation that improvements for MSSSIM are less than those for PSNR. This is because the underlying encoder/decoder networks are trained targeting at PSNR, which learns the mapping relationship between PSNR and conditional tradeoffs. This experiment actually shows some level of flexible of using a different OML metric from the original training metric, so as to tune the reconstruction towards the current need at test time.

\begin{figure}
\centering
\begin{subfigure}[b]{\linewidth}
   \includegraphics[width=1\linewidth]{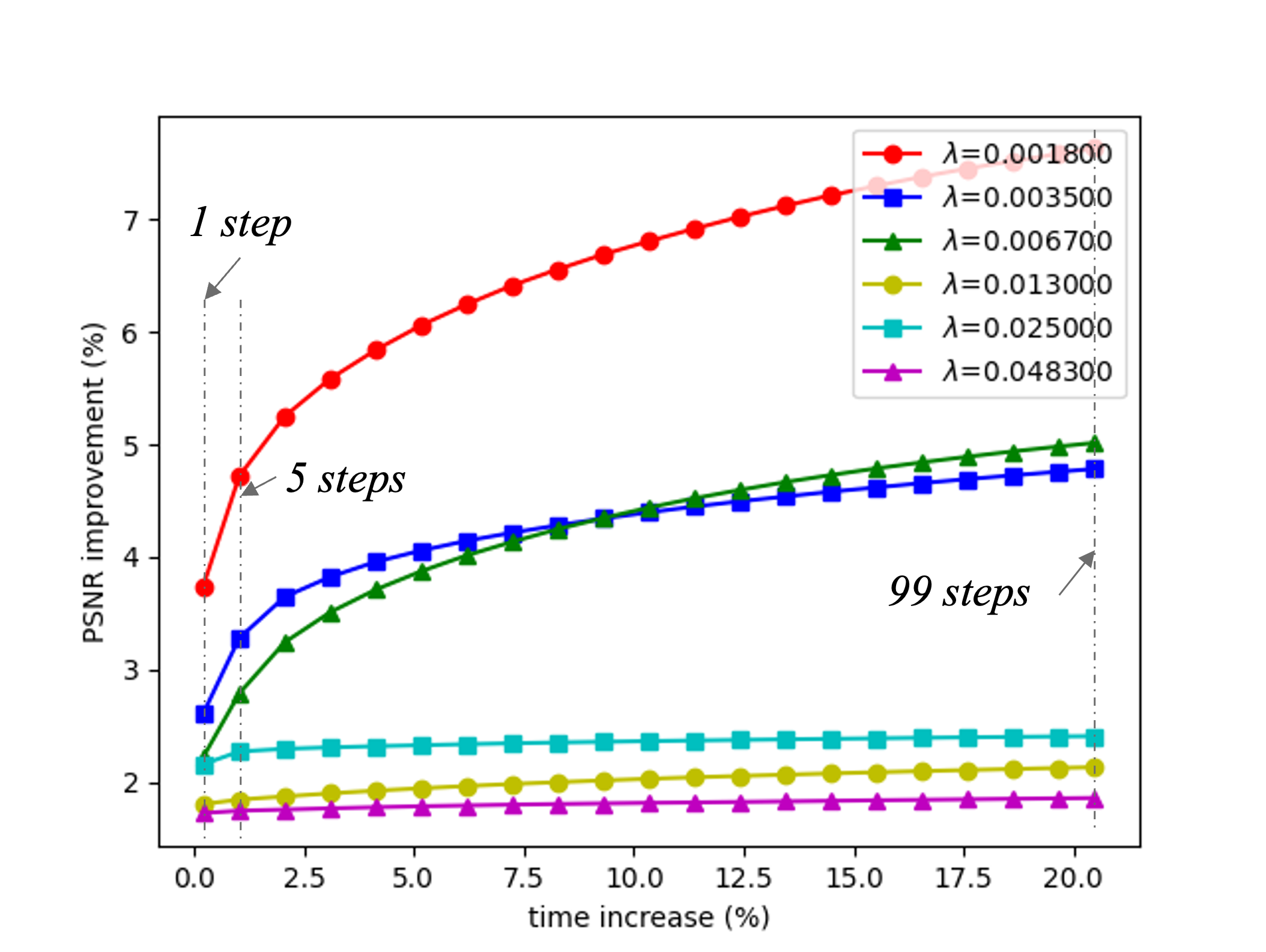}\vspace{-.5em}
\end{subfigure}\vspace{-.5em}
 \caption{Performance vs. time for different iterations.}\label{fig:time}  \vspace{-2em}
\end{figure}

Figure~\ref{fig:time} shows the relationship between the increase of time complexity and the improvement of reconstruction, with different online update steps (iterations). More steps lead to better reconstruction, with a cost of more computation. Since our OML works on decoder network using quantized latent and avoids multiple passes of context modeling, the online iteration is quite fast, \textit{e.g.}, with about $1\%$ encoding time increase for 5 iterations to get $1.5\%\sim5\%$ PSNR gains. With about $20\%$ time increase with 99 iterations, we get $2\%\sim8.8\%$ gains. The decoder still has only one-time inference, and the decoding time remains unchanged.

\begin{figure*}
\centering
\begin{subfigure}[b]{\linewidth}
   \includegraphics[width=1\linewidth]{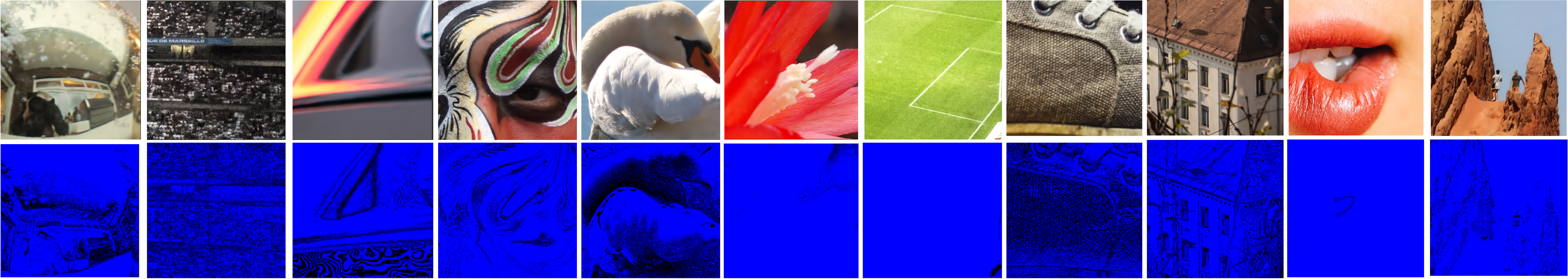}
   \caption{Patches with large gains. The second row shows the changes OML added to the reconstructed images. Brighter color corresponds to larger change.}
   \label{fig:goodpatches} 
\end{subfigure}
\begin{subfigure}[b]{\linewidth}
   \includegraphics[width=1\linewidth]{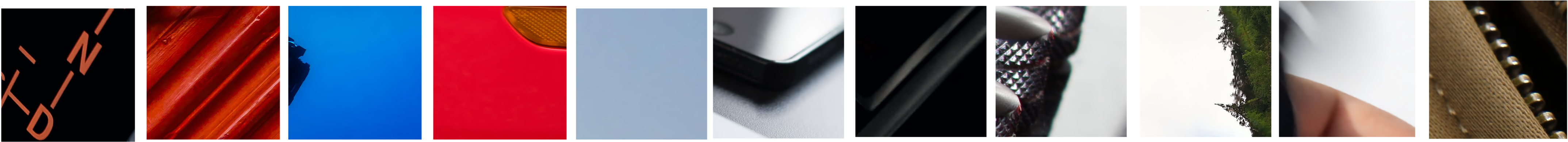}
   \caption{Patches not benefit from online adaptation.}
   \label{fig:nogainpatches}
\end{subfigure}\vspace{-1em}
   \caption{Example patches that do or do not benefit from our OML approach. }\label{fig:patchexample}\vspace{-1em}
\end{figure*}

Figure~\ref{fig:goodpatches} gives some examples of the $512\!\times\!512$ inference patches with large gains from online adaptation for Cheng2020. For PSNR, all patches benefit from online adaptation. While for MSSSIM, OML does not help over some patches, and Figure~\ref{fig:nogainpatches} gives some examples of such patches. Overall, patches with large gains have rich fine details or simple homogeneous textures with visible reconstruction artifacts (blocking effect, noises, \textit{etc.}). Patches without gains are high quality patches, with simple textures comprising of clean homogeneous regions. 

Intuitively, patches with homogeneous textures have small rate losses in nature, while patches with rich fine details have large rate losses. It is hard for encoder/decoder to reduce their rate losses during training, and therefore the model focuses on other patches whose losses can be effectively reduced. Through online adaptation, we can change reconstruction to better attend such patches at test time, \textit{i.e.}, to bridge the gap between their latent distribution and the trained estimation. On the other hand, high-quality patches are less robust to changes induced by online adaptation. 

\begin{figure}
\centering
\begin{subfigure}[b]{\linewidth}
   \includegraphics[width=1\linewidth]{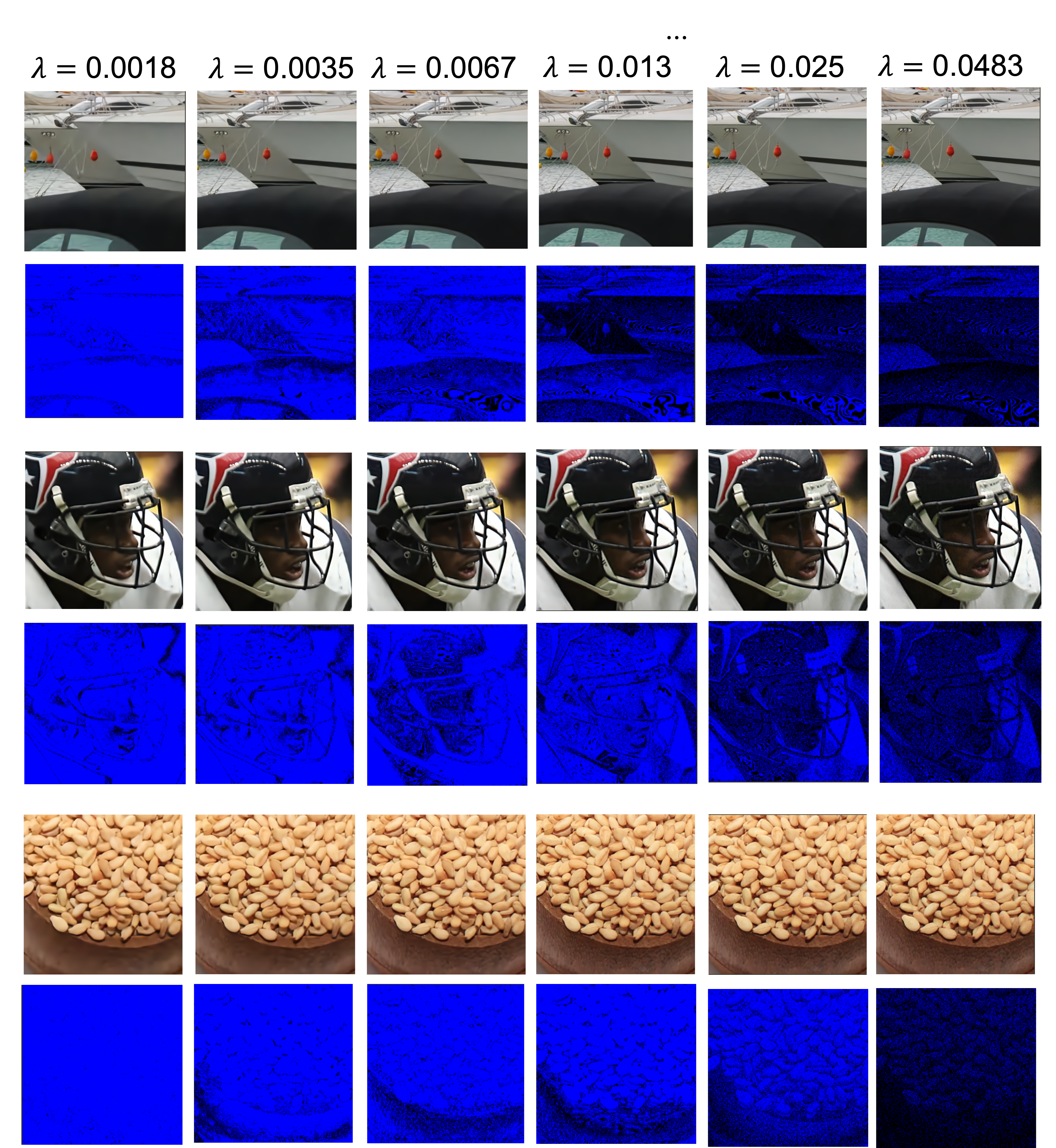}\vspace{-1em}
\end{subfigure}
   \caption{Example of reconstruction and changes varying with $\lambda$.  }\label{fig:comparedetail_lambda}\vspace{-1em}
\end{figure}

Figure~\ref{fig:comparedetail_lambda} further gives some examples of the changes our method makes into the reconstructed patches, as well as how such changes vary with the target tradeoff  $\lambda$. As the bitrate increases, the amount of changes OML makes decrease. For low-quality reconstruction, the OML makes changes all over the place. For high-quality reconstruction, most changes are over the high-intensity regions with uneven textures like water, cloud, blurred content, \textit{etc.} Such results help us to understand the modeling capability of the underlying network in dealing with different content. Improved CVAE networks may be developed to handle such content diversity and bring further performance gains. 


\section{Conclusion}

We proposed an OML framework for LIC using the CVAE architecture. The online learning mechanism is used to bridge the gap between the training soft approximate quantization and the true hard quantization at test time. The meta-learning mechanism is used to control reconstruction with different RD tradeoffs. Direct SGD is used to effectively update the conditional meta parameters so that the decoder can adapt its reconstruction based on the quantized latent representation. Experiments demonstrate the effectiveness of our method. With negligible transmission and computation overhead, our method can boost the performance of different state-of-the-art LIC methods. 

The OML framework also helps to reveal the modeling capability of the underlying LIC network for different image content with different compression qualities. Future work includes further exploring such capabilities and developing content-adaptive CVAE architectures.

\subsection{Limitations}

When the test image has very similar latent distribution to the estimated one, it can not benefit much from our OML. As other LIC methods, we assume a fixed task distribution where the underlying encoder/decoder networks are trained to handle all natural images. Therefore, we do not change the trained network  parameters in OML. This is in contrast to continuous learning where the task distribution can change overtime. In such a case, our OML can be extended to include model adaptation.  

{\small
\newpage
\bibliographystyle{ieee_fullname}
\bibliography{onlinemetaNIC}
}

\end{document}